# I. Introduction

Seed dormancy is a crucial adaptive trait that determines the timing of germination and the successful establishment of seedlings, which are key factors for plant survival and agricultural productivity. It is defined as the temporary inability of a viable seed to germinate under favourable conditions due to physiological or physical constraints [Bewley, 1997]. Dormancy enables plants to synchronise germination with optimal environmental conditions, enhancing the likelihood of seedling survival [Baskin and Baskin, 1998]. Several kinds of dormancy have been described, including physiological dormancy, which is most common in Angiosperms and especially prevalent in temperate seeds [Finch-Savage and Leubner-Metzger, 2006]. Physiological dormancy in seeds can be alleviated through hormonal treatments like gibberellic acid (GA3) which promotes germination by stimulating metabolic processes and breaking inhibitory mechanisms [Finch-Savage and Leubner-Metzger, 2006].

In the Brassicaceae family, dormancy traits show considerable variation. Many species in the family, including wild relatives of cultivated Brassicas, often display significant levels of physiological dormancy that require specific cues, such as cold stratification or hormonal treatments to break this dormancy [Baskin and Baskin, 2004]. Deep dormancy helps seeds persist in the soil for extended periods, enhancing species survival in adverse conditions. However, it can be a barrier to the utilisation of plant species or their wild relatives in agriculture, where rapid and uniform germination is desired [Bewley, 1997].

For the *Brassica* genus, diverse germination phenotypes have been documented, influenced by temperature, water potential and other environmental factors. Studies have shown that thermal- and hydrotimes for germination can vary significantly, with base temperatures ranging from 1.7 to 10.5° C and base water potentials ($\Psi_b$) from -1.5 to -0.1 MPa, indicating considerable adaptive variation within the genus [Castillo-Lorenzo *et al.*, 2019]. This diversity suggests that *Brassica rapa* L. has evolved mechanisms to cope with varying climatic conditions, making it a valuable model for studying germination responses to environmental stress [Aissiou *et al.*, 2018].

This intrinsic variability in germination responses within the *Brassica* genus suggests that dormancy-breaking strategies may need to be tailored to specific environmental and physiological conditions. In this context, novel approaches such as cold plasma treatment offer a promising alternative to conventional methods, particularly for species exhibiting complex dormancy mechanisms. Cold plasma refers to a partially ionised gas in which electrons are at a much higher temperature than the heavy particles (ions and neutrals), resulting in a non-equilibrium state. This allows the gas to remain at a low overall temperature, close to ambient, while still producing a diverse array of reactive species such as ions, radicals and UV photons [Douat *et al.*, 2023]. These reactive species can interact with the seed coat, leading to physical and chemical surface modifications that alter properties like permeability, wettability and surface energy.

More specifically, the pioneer works of Šerá *et al.* (2009) showed that cold plasma treatment can break seed dormancy in *Chenopodium album*, underlining that seed coat modification was induced by plasma etching or erosion, particularly using argon/oxygen or argon/nitrogen gases. Similarly, plasma treatment overcame dormancy in *Pityrocarpa moniliformis*, raising germination from ~12 to ~38%, a result attributed to increased seed coat wettability and faster water uptake [Nicolau *et al.*, 2022]. Comparable effects were observed in *Medicago sativa*, where both direct plasma and plasma-activated water treatments boosted germination rates and early seedling vigour [Jirešová *et al.*, 2021]. In *Leucaena* seeds, plasma exposure enhanced hydrophilicity, facilitating new water entry pathways that allowed water to penetrate previously impermeable layers, thereby partially overcoming dormancy [Alves-Junior *et al.*, 2020]. While these findings highlight plasma-induced dormancy-breaking mechanisms (primarily through enhanced water absorption and seed activation processes), it is also worth stressing that cold plasma offers a sustainable and chemical-free alternative to traditional dormancy-breaking approaches, such as gibberellic acid (GA3) treatments and mechanical scarification [Dufour *et al.*, 2021], [August *et al.*, 2023].

Despite these advances, the effectiveness of cold plasma treatment compared to conventional methods remains underexplored. The BRASEXPLOR project, supported by the PRIMA European H2020 initiative, collected 63 wild populations and 68 landraces of *B. rapa* from various Mediterranean regions to investigate germination patterns and other morphological and functional traits [Falentin *et al.*, 2024]. To obtain homogeneous experimental material for the project purposes, all seeds were produced under standard conditions at Le Rheu, France. Some populations showed very low germination even when cold stratification was applied. This has led to the search for an effective dormancy breaking/ germination protocol to sustain propagation in *ex situ* genebanks. This project highlighted the variability in dormancy patterns across populations, particularly in those from arid environments such as Italian and Algerian regions which showed poor germination. These findings underlined the importance of evaluating multiple approaches to release dormancy and optimise germination for conservation and agricultural applications.

Seed coat-imposed dormancy in *B. rapa* has not been widely documented, although one study has reported enhanced dormancy associated with seed coat properties in *Sinapis arvensis* L., a related species [Duran and Tortosa, 1985]. Given the variability in dormancy mechanisms across populations, the current study aims to provide a comprehensive analysis of the dormancy-breaking efficacy of cold plasma, GA3, and scarification treatments. By comparing these methods across 61 wild *B. rapa* populations from different Mediterranean regions, this research seeks to unravel the complex interplay between seed coat structure, dormancy level, and environmental adaptation. Our results will not only enhance our understanding of seed dormancy mechanisms in wild *Brassica* species but also inform sustainable approaches for *ex situ* conservation.

## II. Materials and methods

### II.1. Seed material and environmental traits

Seeds from wild *B. rapa* populations were produced in Le Rheu, France in 2021/22 after collecting seeds from at least 15 wild mother plants from native areas all around the Mediterranean basin (**Figure 1**). To ensure a minimum of 30 seedlings per population, 60 seeds per population, collected from all mother plants, were sown. Ten plants per population were selected for vernalisation (one month at 4-6°C, eight-hour day length and watering for 4-5 leaf-stage seedlings). Then, plants were transplanted under cages for flowering and seed production: five plants per cage and a total of two cages (10 plants) per population. Flies were added under each cage to promote pollination. At maturity, all cages of the same population were harvested at the same time and seeds were mixed all together after threshing to get one common batch per population for all experiments [Falentin *et al.*, 2024]. Seed lots were stored for approximatively 18 months at 4°C and 10% relative humidity in the dark until phenotyping, when lots were split for different analyses in controlled conditions or emergence in greenhouse experiments.

The original, geo-referenced locations of the collected populations was used to extract climate data from WorldClim with an accuracy of 1 km for data over the period 2010-2018. France has the highest average annual precipitation (83.7 mm), Algeria has a moderate average precipitation (68.6 mm) with significant variability and Sicily, Italy, the lowest precipitation (54.1 mm), reflecting typical Mediterranean climates (**Figure 1**). The native area of Brassica is therefore large and interesting for collecting seed material to study local adaptations to wet, arid or semi-arid environments. Regarding temperature, Algerian populations experience the highest mean maximum temperatures (31.5°C), indicating extreme heat, particularly in inland regions. French populations live in cooler conditions, with the lowest mean maximum temperatures (24.8°C) and occasional frosts. Sicily shows intermediate temperatures (mean maximum 28.4°C), typical of Mediterranean climates, with mild winters and warm summers.

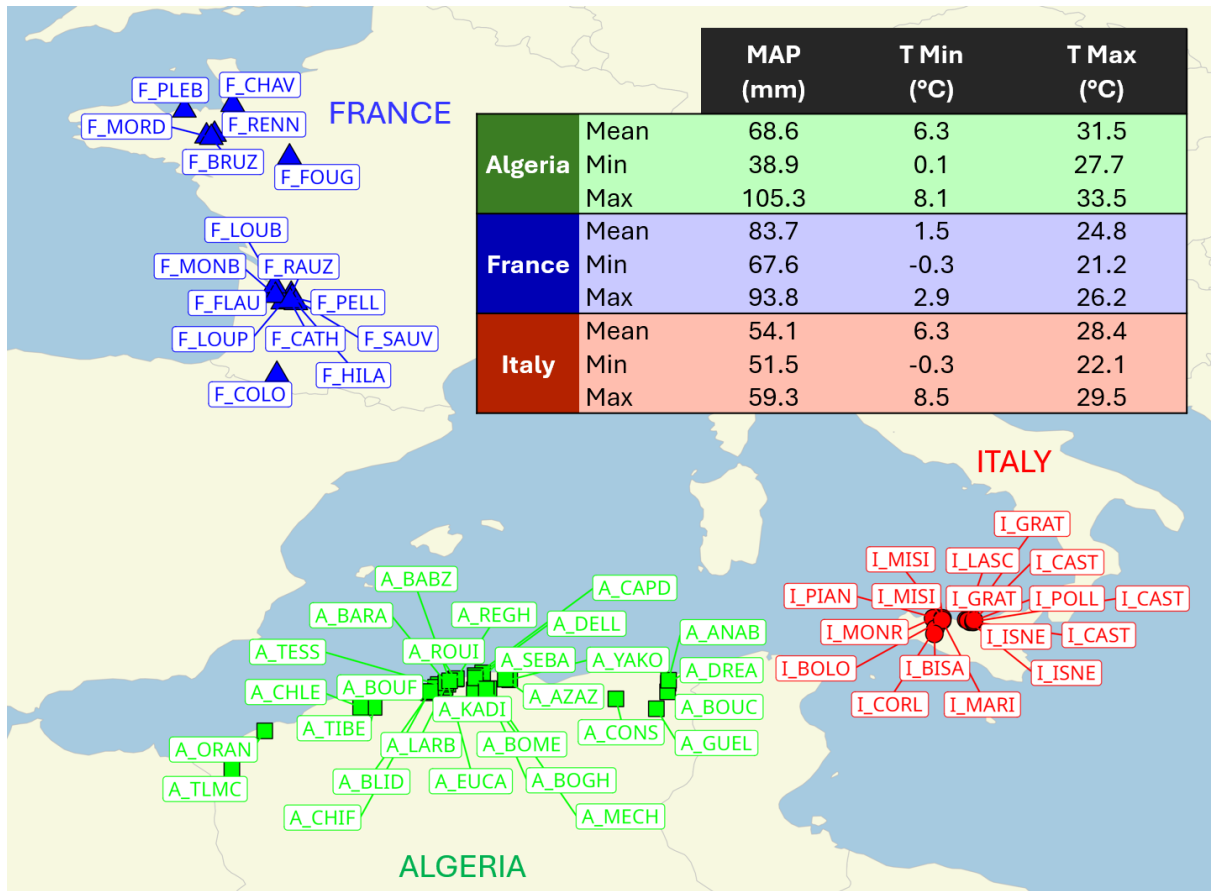

*Figure 1. Geographical origin and climate variables corresponding to the collection sites of 61 wild populations of Brassica rapa from around the Mediterranean basin. Parameters were extracted from Worldclim database during the period 2010-2018. MAP = mean annual precipitation (mm), $T_{Min}$ = minimum temperature (°C) and $T_{Max}$ = maximum temperature (°C).*

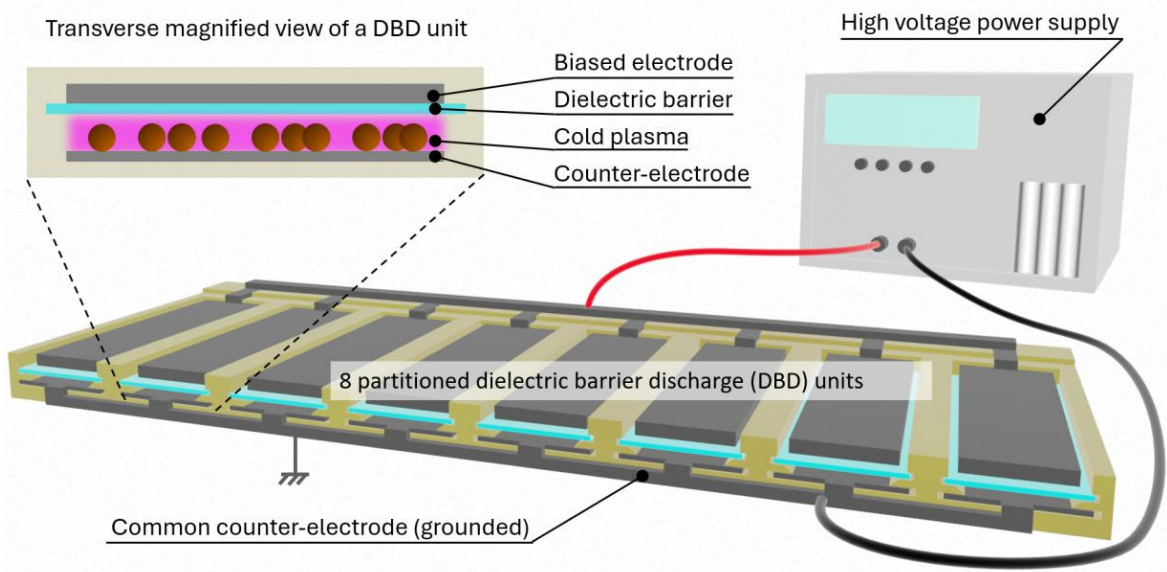

*Figure 2. Diagram of the experimental setup for cold plasma treatment of Brassica rapa samples (1050 seeds per population). The device is partitioned into eight individual DBD's where cold plasma of ambient air is generated.*

## II.2. Methods for releasing seed dormancy / Cold plasma treatment

The 61 *B. rapa* seed populations were treated using a cold plasma of ambient air generated in a system of eight partitioned dielectric barrier discharge (DBD) units, designed to operate in parallel. This setup allowed for the simultaneous plasma treatment of eight distinct seed populations, each containing 1050 seeds, optimising the throughput and efficiency of the treatment process (**Figure 2**).

Each DBD unit was configured in a plane-to-plane arrangement, consisting of a 1 mm thick dielectric barrier and two electrodes: a high-voltage electrode (stainless steel, 20 × 60 mm$^2$) and a grounded counter-electrode (stainless steel, 26 × 80 mm$^2$). The gap distance between the dielectric barrier and the counter-electrode was fixed to 2 mm for all experiments. The high-voltage electrode was powered using a high-voltage generator, which included a function generator (ELC Annecy, France, GF467AF) and a power amplifier (Crest Audio, 5500W, CC5500). A ballast resistor (250 kΩ, 600 W) was incorporated into the power supply circuit to limit the current flow. The applied high voltage was sinusoidal, 6 kV in amplitude at 500 Hz. The gap between the dielectric barrier and the grounded electrode was set to 1 mm.

Electrical monitoring was achieved using a digital oscilloscope (Teledyne LeCroy Wavesurfer 3054), featuring a 500 MHz bandwidth and a 4 GS per second sampling rate. The oscilloscope was configured to measure three channels: the voltage at the biased electrode and counter-electrode, monitored using high-voltage probes (Tektronix P6015A, 1000:1; Teledyne LeCroy PPE 20 kV, 1000:1) connected to channels CH1 and CH2, respectively. The discharge current was measured at CH3 using a current transformer (Pearson model 2877), that could be positioned between each of the eight counter-electrodes and the ground.

In this study, cold plasma was generated using a sinusoidal voltage of 6 kV amplitude at 500 Hz, applied for 10 minutes as a standard, non-optimised treatment for seed processing.

## II.3. Germination monitoring

To monitor seed germination of 61 populations, germination percentage was scored every two hours for seven days on four replicates of 25 seeds sown on automated germination tools regulated at 20 ± 0.5°C for control and plasma treated samples. This in-house equipment allows accessing dynamic germination time courses through digital imaging [Demilly *et al*., 2014]. Germination rate was estimated considering two parameters: the time to 10% germination ($T_{10}$) and the mean germination time (MGT), both calculated on day-5 to minimise false positives caused by elongated radicles displacing ungerminated seeds in the later stages. MGT corresponds to the average of the germination times calculated individually for each seed.

MGT averages the germination times of all seeds that do germinate. But when total germination is very low (e.g., 5%), MGT reflects only a very selective and possibly unrepresentative subgroup. In contrast, when germination is high (e.g., 60%), MGT reflects a much larger and more representative portion of the population. Since this inconsistency could make comparisons misleading, $T_{10}$ was also measured as a fixed benchmark. $T_{10}$ always refers to the time when 10% of the sown seeds have germinated so that comparison is based on the same threshold. This makes it more robust and comparable, especially across accessions with widely different dormancy levels. Germination percentage, $T_{10}$ and MGT were jointly analysed to characterise both the extent and dynamics of seed germination.

At the end of the 7-day experiment, all non-germinated seeds of 15 populations that had achieved only 20 to 30% germination (five from each country) were transferred to Petri dishes with filter paper moistened with 0.2% GA3 (Sigma Aldrich) aqueous solution. They were subsequently incubated at 20°C to test germination for seven more days. The remaining non-germinated seeds were then mechanically scarified one by one and left for another 7-day period at 20°C.

## II.4. Dormancy monitoring

A second experimental protocol was set up to evaluate the type of dormancy and the effect of plasma on breaking seed dormancy. Four replicates of 25 seeds per population were sown in Petri dishes (diameter 9 cm), on two layers of filter paper soaked in 6 mL of either distilled water (for control, plasma and scarification tests) or 0.2% GA3 aqueous solution. All seeds were rinsed in distilled water and blotted dry before sowing. For chemical scarification, 25 seeds per population were placed in a 15 mL test tube and 2 mL of 5% NaOCl commercial bleach was added to each tube, ensuring that the seeds were fully submerged. The seeds were gently agitated by swirling the tubes every 2-3 minutes during a 10-minute treatment, to ensure even exposure to the bleach. After scarification, the bleach solution was carefully decanted, and the seeds were immediately rinsed twice with 10 mL of distilled water to remove residual NaOCl. Seeds were re-dried on filter paper and sown immediately in Petri dishes. Petri dishes were incubated in the dark at 20°C for seven days. During this period, every 24-hours germinated seeds were scored manually and removed from the dishes. Scoring was conducted in a dark room under low intensity green light.

## II.5. Greenhouse emergence

For each of the 61 populations, 100 untreated control seeds and 100 plasma-treated seeds were used. On 5 June 2023, between 25 and 30 seeds were sown in standard potting soil (Faliénor substrate with Ferti) in a greenhouse (25 ± 5°C) at Le Rheu in a randomized complete three block design. The emergence (number of seedlings) was scored two weeks after sowing.

## II.6. Scanning electron microscopy (SEM) and histological studies of seed coat

Six populations were used for SEM and histochemical studies of seed coat. Five mature seeds per population were used for micro-sculpture observations. SEM was performed using JEOL JSM-7610FPlus (Tokyo, Japan) a Schottky Field Emission (S-FEG) scanning electron microscope.

Three to five mature seeds per population were used for anatomical and histochemical analysis. The detection of lignins was performed on fresh cross-sections using the HCl phloroglucinol reaction according to Zhou *et al.* (2009). Hand serial sections were made using a razor blade. The fresh sections were soaked for three minutes in 1% phloroglucinol solution and then mounted in a drop of concentrated hydrochloric acid. The phloroglucinol-HCl reagent reveals the aromatic structures of aldehydes contained in lignins by producing a red staining. The unsubstituted cinnamaldehyde functional group is responsible for this colouration. This technique allows for quick visualisation of lignified tissue walls containing cinnamaldehydes, which are precursors of lignin. The immediately stained sections were observed under a Leica DM750 microscope with 4×, 10× and 40× objective lenses, and images were captured with a digital camera (Leica ICC50-E).

# III. Results

## III.1. Seed dormancy and germination spread according to the seed native region

The 61 wild Brassica rapa populations, collected from diverse environments, were grouped by geographic origin to identify overall trends in dormancy and germination.

Seed lots from Algerian populations showed a broad range of $T_{10}$ values, from approximately 35 to 110 hours, and germination percentage between 5 and 60%, reflecting significant variability in dormancy traits (**Figure 3**). Seed lots from French populations typically had lower $T_{10}$ values, mostly between 25 and 70 hours, and higher germination percentages (up to 80%). The $T_{10}$ for seeds derived from the Italian populations was between 30 and 120 hours, with germination percentages between 5 and 50%, suggesting moderate germination dynamics. For each country, populations with lower $T_{10}$ values (faster germination) showed generally higher final germination percentages, especially for populations native from France which were less dormant, a significant negative correlation was obtained between $T_{10}$ and germination percentage (r = –0.89; –0.93 for control samples and 0.80 for plasma treated samples). Conversely, populations with longer $T_{10}$ values generally exhibited lower germination percentages. Despite being a second generation of seeds produced in temperate conditions, dormancy levels were similar to the levels observed for the original seeds harvested in native conditions [Wagner et al., 2023]. Finally, no clear relationship between seedling emergence (bubble size) and the other germination parameters ($T_{10}$ and germination percentage) was found, indicating no distinct trends in emergence across the different populations.

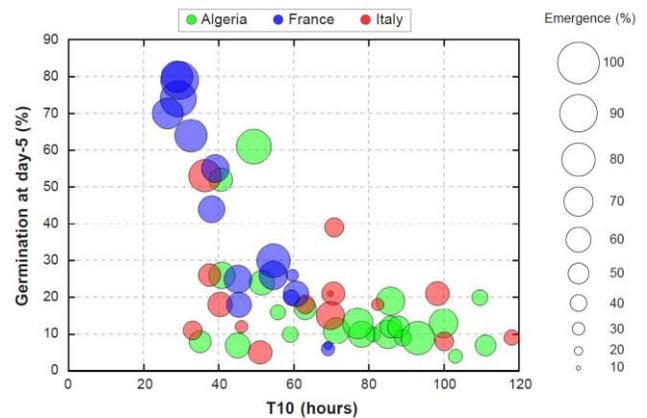

*Figure 3. Bubble chart illustrating the relation between time to reach 10% germination ($T_{10}$) and germination percentage at day-5 for untreated Brassica rapa seeds from 61 populations. Bubble size indicates emergence percentage, colour-coded by country of population origin (Algeria, France, Italy).*

## III.2. Assessing cold plasma efficacy in breaking seed dormancy and accelerating germination

In this study, seeds representing all the 61 B. rapa populations were exposed to a cold plasma of ambient air for 10 minutes, at 6 kV and 500 Hz. The efficacy of this standard treatment in seed dormancy release was assessed through germination percentage, $T_{10}$, MGT and emergence percentage.

Plasma treatment significantly increased the germination percentage, to a mean of approximately 60% across all populations, compared to 18% for the control group (**Figure 4a**). The broader interquartile range in the plasma-treated group indicates greater variability in germination responses among populations. Several outliers in the control group reveal that, while most populations show low germination percentages without treatment, some exhibit relatively high germination. Their germination was not affected after cold plasma exposure proving that this seed treatment was noninvasive. **Figure 4b** further illustrates this, showing control populations sorted from the least dormant (80% germination) on the left to the most dormant (as low as 3% germination) on the right. Plasma-treated seeds generally achieved higher germination percentages, particularly in more dormant populations. For example, populations such as A-TIBE and I-ISNE showed marked increases, from 5% in control to over 55% with plasma treatment. In contrast, populations like F-PLEB and A-TLMC, which had germination percentages above 50% under control conditions, exhibited only modest increases of 10% or less following plasma exposure.

Plasma treatment appeared to accelerate the germination process, reducing $T_{10}$ from 61 hours to 37 hours, and MGT from 66 hours to 56 hours (**Figure 4c, 4e**). The larger reduction in $T_{10}$ compared to MGT suggests that plasma treatment had a stronger effect on the initiation of germination than on the overall average speed. In other words, the treatment mainly benefited the fastest-responding seeds, triggering earlier germination in a subset of the population. While MGT also decreases, the smaller gap indicates that later-germinating seeds were less affected. This implies that plasma primarily acts by reducing early dormancy barriers, rather than uniformly accelerating germination across all seeds. In **Figure 4d**, where the control populations are ranked from the longest $T_{10}$ (120 hours) to the shortest (27 hours), it appears that plasma treatment reduces $T_{10}$ across all populations except A-YACO. Considering **Figure 4f**, it appears that cold plasma significantly decreased MGT except for nine seed lots (e.g., I-PIAN, I-BISA and I-ISNE). Overall, it appears that the effectiveness of plasma treatment in reducing $T_{10}$ and MGT did not appear to depend on their initial values.

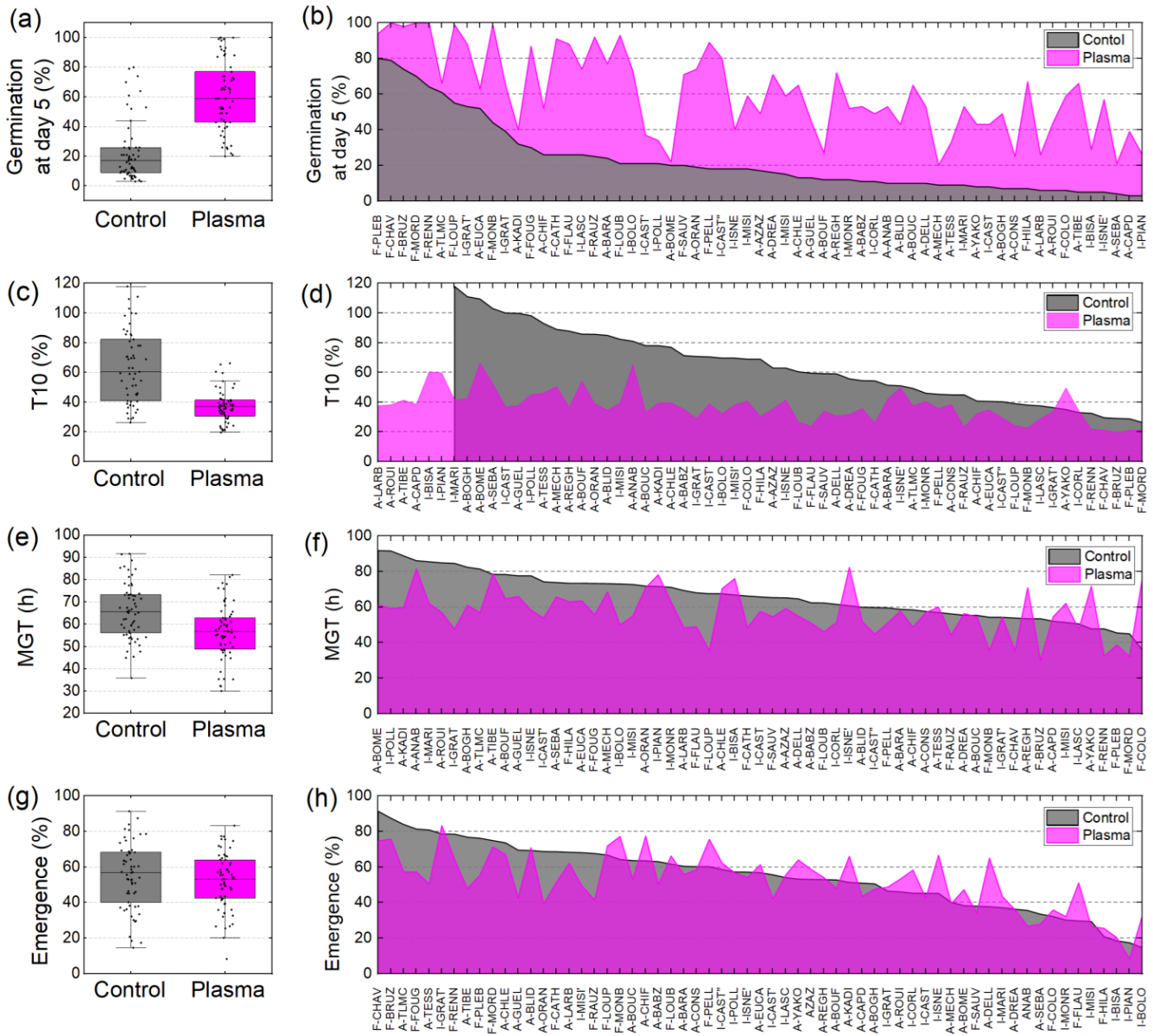

*Figure 4. (a) Box plot comparing the 5-day germination percentage of the 61 Brassica rapa populations under untreated (Control) and plasma-treated (Plasma). (b) Germination percentage at day-5, arranged from the least to the most dormant populations. (c) Box-plot comparing the time to reach 10% germination between control and plasma-treated groups. (d) Time to reach 10% of germination for untreated and plasma-treated seeds across the 61 Brassica rapa populations, sorted from longer to shorter. (e) Box-plot comparing the mean germination time between control and plasma-treated groups.; (f) Mean germination time for untreated and plasma-treated seeds across the 61 Brassica rapa populations, sorted from longer to shorter MGT. (g) Box-plot comparing seedling emergence between control and plasma groups. (h) Seedling emergence of untreated and plasma-treated populations, sorted from higher to lower emergence values.*

Finally, emergence on day-15 was comparable between control and plasma-treated groups, both averaging around 55% (**Figure 4g**). Although the range of emergence values was slightly narrower for plasma-treated populations, the substantial overlap suggests that plasma treatment had a limited impact on enhancing emergence, especially in populations with higher baseline emergence. Ordering control populations from the highest (> 80%) to the lowest (< 10%) emergence percentage reveals that plasma treatment showed limited effectiveness for populations with baseline emergence above 70% (**Figure 4h**). In some cases, such as F-RAUZ and A-TESS, plasma treatment even appeared to have a detrimental effect on populations with high initial emergence.

An increase in emergence percentage by 10% or more compared to the control was noted in only 4 of 26 Algerian, 3 of 16 French, and 3 of 17 Italian populations. Conversely, plasma treatment adversely affected some populations, resulting in an average 12% decrease in emergence; this occurred in 6 of 26 Algerian, 7 of 16 French, and 2 of 17 Italian populations.

## III.4. Efficacy of cold plasma treatment in releasing deep dormancy

Since cold plasma significantly improved germination in the 61 B. rapa populations across the three countries and significantly reduced MGT in most of them, we evaluated whether the treatment was more effective at alleviating embryo or seed coat-imposed dormancy. To address this, we compared the effects of plasma treatment with conventional methods: scarification and GA3 application, which specifically target these different dormancy levels (**Figure 5**). Seven days after imbibition using Petri dishes test, germination reached 67% with cold plasma, 71% with GA3 and 79% with chemical scarification, compared to only 48% for untreated seeds (vertical bars). The interquartile range (IQR) for the control was 33% (65-22%), indicating substantial variability. Plasma-treated seeds showed a narrower IQR of 26% (83-57%), suggesting reduced variability, though some heterogeneity remained, as indicated by outliers. GA3 treatment had a wider IQR of 40% (89-49%), reflecting more heterogeneous responses despite a higher germination percentage compared to plasma treatment. Scarification, with a relatively narrow IQR of 21%, demonstrated greater consistency in its effects, although outliers were still present.

The efficacy of cold plasma treatment in releasing dormancy of the 61 wild B. rapa populations was therefore significant and performed comparably to GA3 while offering better homogeneity. Nevertheless, scarification provided the most consistent results, regardless of the statistical measure (mean, median, IQR).

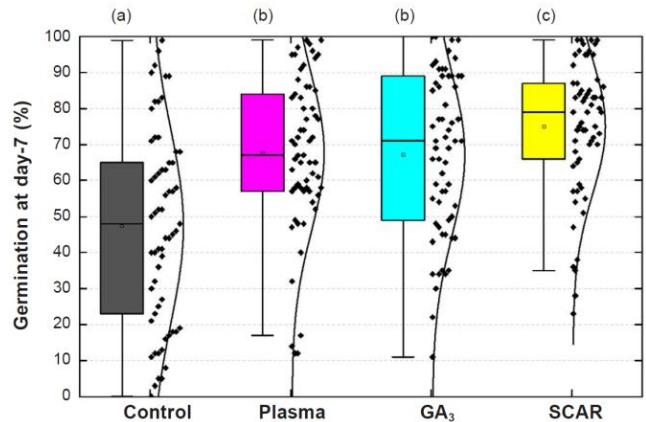

*Figure 5. Comparisons of the day-7 germination for seeds from 61 Brassica rapa populations considering four treatments: control, cold plasma, GA3 and scarification with bleach (SCAR). Box plots display mean values, interquartile range and outliers, while density distributions illustrate germination percentage variability across treatments. Horizontal line in the box is the median. Open white square in box is the mean. Lowercase letters above the box plots indicate statistically homogeneous groups based on Tukey's HSD test applied to mean germination percentages. Treatments sharing the same letter are not significantly different (p > 0.05).*

## III.5. Exploring physiological dormancy depth using GA3 and chemical scarification

A classification of the 61 B. rapa populations based on germination percentage differences following GA3 and scarification treatments relative to control is reported in **Figure 6**. Dormant populations, i.e. with germination below 50% were identified for Algeria, France and Italy. Non-dormant populations with native germination higher than 50% were identified for Algeria and France (blackened symbols). All datapoints were distributed across different sections of the graph, each representing distinct dormancy characteristics and treatment effectiveness:

- In the dashed box section, populations exhibited less than 10% difference of germination between treated and control groups, indicating two possible scenarios: (i) the seeds were dormant but neither GA3 nor scarification significantly alleviated dormancy; or more likely, (ii) the control seeds had high native germination percentages (e.g. > 50%), indicating a lack of dormancy and therefore the irrelevance of using scarification or GA3 treatments.
- In section I, germination was lower for both scarification and GA3 treatments compared to the control, suggesting that these treatments were either detrimental or ineffective in promoting germination, indicating a potential lack of physiological seed maturity.
- In section II, GA3 enhanced germination relative to the control, while scarification reduced it, suggesting single physiological dormancy, selectively released by GA3. Only F-RENN exhibited physiological dormancy, although the effect was minimal because the datapoint was at the threshold of the dashed section (germination percentage > 50%).
- Section III highlights the scenario where scarification significantly increased germination percentage compared to the control, while GA3 did not. This indicates that scarification effectively broke dormancy. Examples include Algerian populations (A-TIBE, A-REGH, A-DELL), which displayed deep dormancy, albeit weakly alleviated by scarification (around 10%).

- Section IV is where both scarification and GA3 improved germination compared to the control. This section is divided into two sub-sections denoted IVa and IVb. In subsection IVa, scarification was more effective than GA3, indicating that dormancy was associated with the seed coat. Many Italian populations, such as I-GRAT, I-CAST and I-ISNE, are located here. In sub-section IVb, GA3 treatment resulted in higher germination compared to scarification, suggesting a predominance of embryo dormancy. Populations like F-HILA and I-BISA showed this kind of dormancy.

The scarification and GA3 treatments did not significantly enhance germination in 7 out of 61 populations (those within the dashed section).

Complementary trials carried out after germination monitoring for five populations per country having less than 35% of germination, after seven days at 20°C confirmed that embryo and seed coat could both impose dormancy in the same population (**Figure 7**). Germination could reach 20 to 50% when seeds were transferred upon paper moistened with a solution of GA3 but was fully achieved after single seed mechanical scarification. For A-TIBE, only scarification was efficient, so seed coat-imposed dormancy could be confirmed for this population. For French populations, embryo dormancy was the main type but three of them contained around 20% of deep dormant seeds.

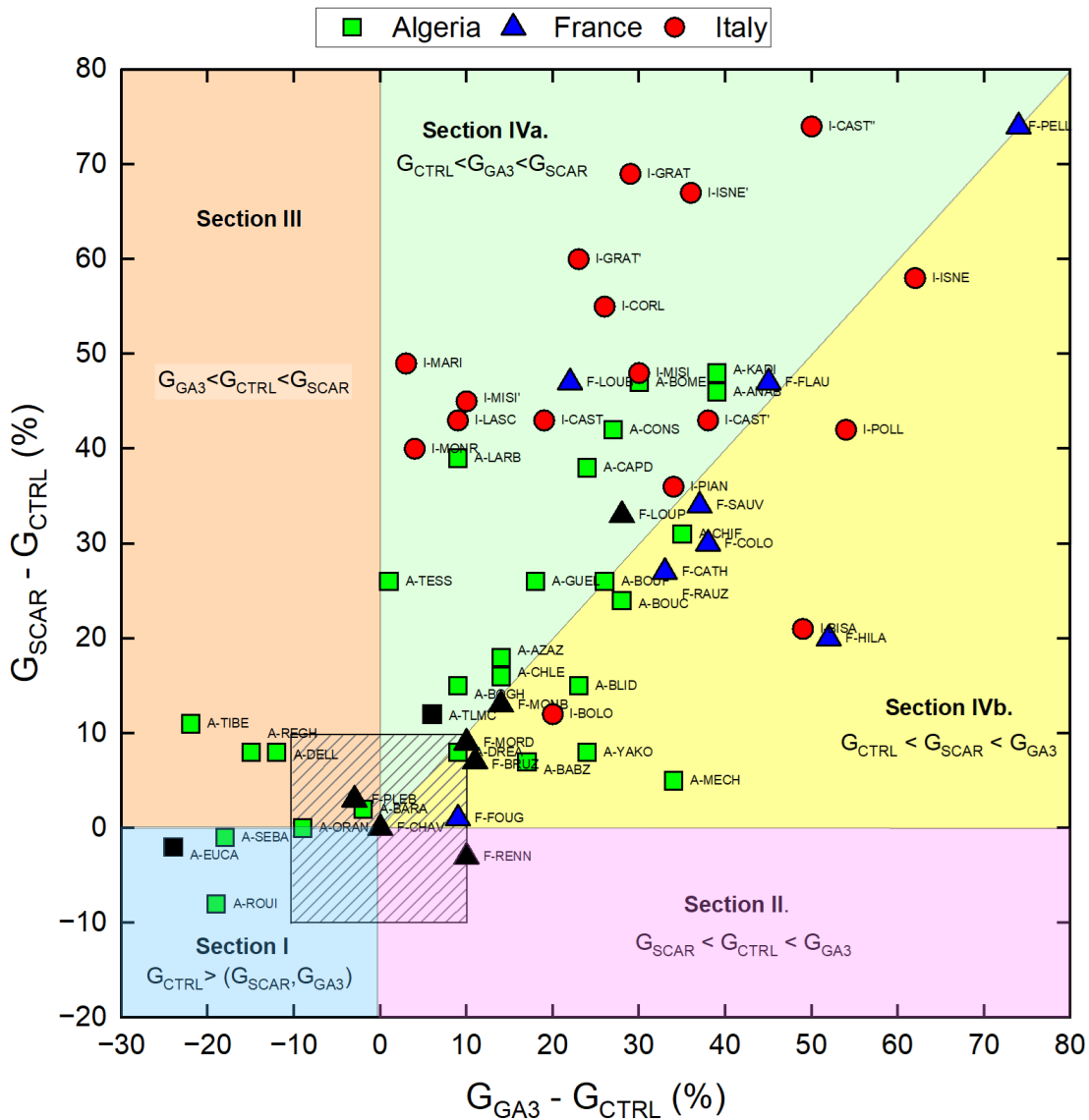

*Figure 6. Classification of the 61 Brassica rapa populations based on germination percentage differences following GA3 and scarification treatments relative to control. Five regions are indicated based on dormancy release efficiency: non-dormant (CTRL >), hormonal release (GA3 >), scarification release (SCAR >), predominantly scarification release (IVa) and predominantly hormonal release (IVb). The dashed region is where SCAR and GA3 treatments do not significantly release dormancy. Symbols are shaded black for populations with native germination percentages above 50%, indicating limited dormancy.*

## III.6. Microscopy observations

To investigate the structural basis of seed dormancy, six Brassica rapa populations were selected for scanning electron microscopy (SEM) based on their geographic origin and predominant dormancy levels. Populations A-ANAB (Algeria) and I-BISA (Italy) exhibited intermediate dormancy, while A-TIBE (Algeria) and I-MISI (Italy) were characterised by a deep dormancy. In contrast, the French populations F-HILA and F-RAUZ predominantly exhibited non-deep dormancy (**Figure 8, Figure 9**).

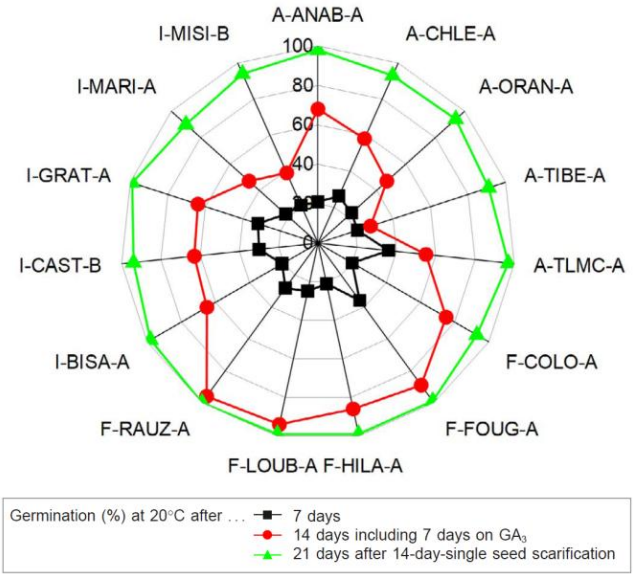

*Figure 7. Average germination (%) after 21 days at 20°C for 15 wild populations (5 per country). Seeds were first sown on automated germination tools; non-germinated seeds after seven days were transplanted onto filter paper moistened with 0.2% of GA3 for seven days and then mechanically scarified individually before another 7-day period.*

SEM analysis revealed that the Algerian and Italian seeds had a distinct reticulate seed coat architecture, marked by prominent, polygonal anticlinal cell walls and irregularly undulating periclinal surfaces-morphological features (**Figure 8**). Conversely, the French populations displayed a reticulate-foveolate pattern, with thinner, less elevated anticlinal walls and small foveolae dispersed across the periclinal surface, potentially facilitating water permeability and correlating with a non-deep physiological dormancy.

No structural damage or disruption of the seed coat was observed following plasma treatment. Although overall surface morphology remained unchanged, a slight smoothing and reduction in the undulations of the periclinal walls was noted in some Italian seeds, suggesting minimal but detectable surface modification.

To complement the SEM observations of seed coat morphology, optical microscopy was used to examine internal seed coat composition, specifically lignin distribution, which plays a crucial role in determining impermeability and dormancy depth. The Algerian and Italian seed coat exhibited a more intense palisade cell walls staining indicating greater lignin impregnation and a thicker, uniformly lignified palisade layer with closely packed cells, suggesting strong physical dormancy due to enhanced impermeability. In contrast, the French population displayed a thinner, less uniform lignin-rich layer with more loosely arranged cells, indicating a reduced physical barrier (**Figure 9**).

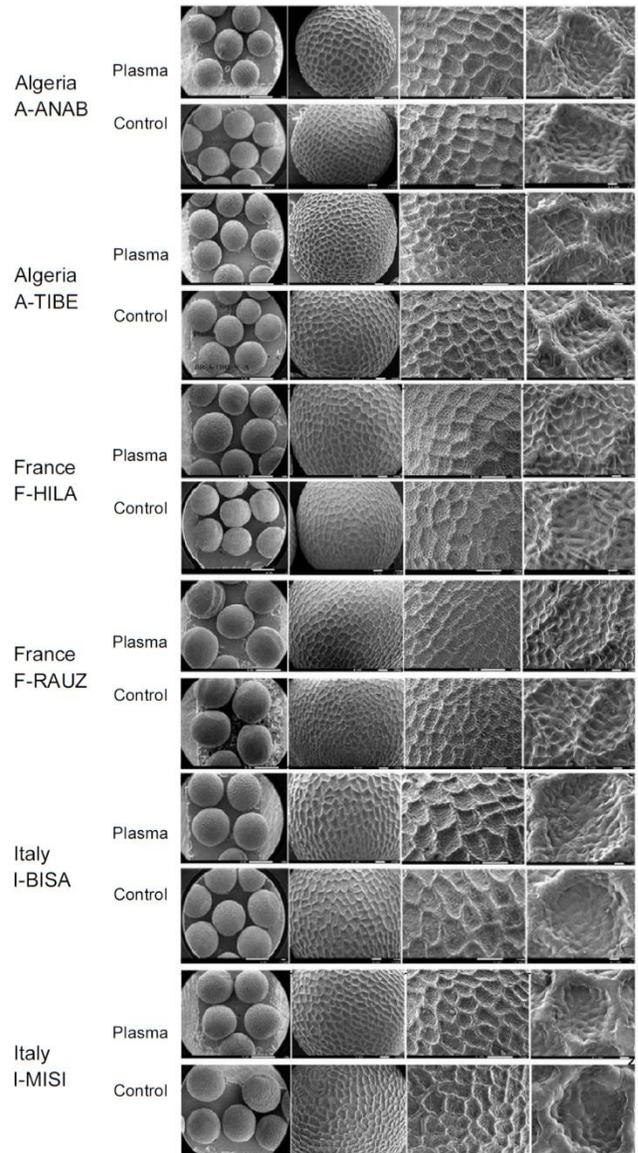

*Figure 8. SEM images showing the surface morphology of Brassica rapa seed coats from six populations (Algeria, A-ANAB and A-TIBE; France, F-HILA and F-RAUZ; Italy, I-BISA and I-MISI) under control and plasma treatments. Each row compares untreated (Control) and plasma-treated (Plasma) seeds at increasing magnifications.*

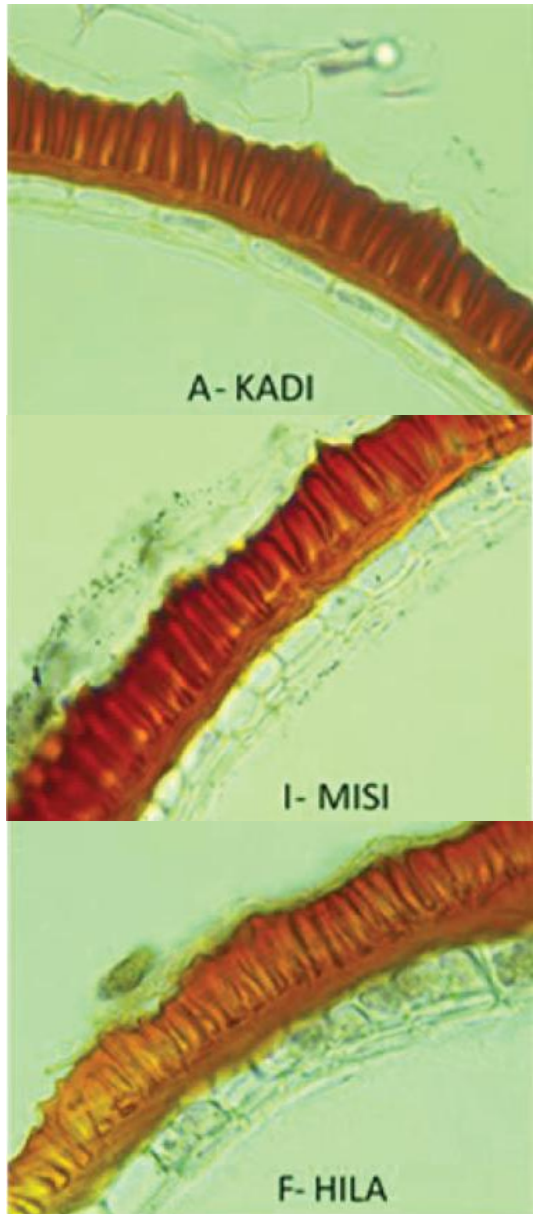

*Figure 9. Phloroglucinol-stained transverse sections of Brassica rapa seed coats, comparing the Algerian population (A-KADI), the Italian population (I-MISI) and the French population (F-HILA). The staining highlights lignin-rich layers, revealing differences in lignification and cell structure between the three populations.*

## IV. Discussion

A combination of genetic and environmental factors influences the complex traits of seed dormancy and germination. The physical structure of the layers surrounding the embryo and the hormonal regulation of its growth, are key factors in breaking dormancy [Koorneef *et al*., 2002]. For wild populations, dormancy is a key ecological factor that guarantees survival under diverse environmental conditions and prolonged persistence in the soil seed bank [Finch-Savage and Leubner Metzger, 2006]. Seed dormancy has been scarcely described in *B. rapa*. The Brassica genus was found to have diverse germination phenotypes, allowing to be the source of adaptive traits for future agriculture [Castillo-Lorenzo *et al*., 2019] and for many related species, physiological dormancy was prevalent [Baskin and Baskin, 2004] and in some cases, a photoinhibition of germination in cool conditions [Manalil *et al*., 2018]. The Sicilian and some Algerian populations exhibited a similar type of dormancy enhanced by seed coat, possibly associated with their arid-Mediterranean habitat. This dormancy has been conserved after seed production in a different environment from native area [Tiret *et al*., 2025]. Differences in lignification may reflect distinct dormancy strategies, with the Algerian and Italian seeds relying on deeper dormancy, while the French seeds potentially favour a more flexible physiological dormancy to respond to moisture conditions.

The results of this study reveal significant insights into seed dormancy mechanisms in wild *B. rapa* populations and the variability in their responses to dormancy release treatments, including cold plasma, gibberellic acid and scarification. More specifically, they highlight the complexity of dormancy regulation, shaped by genetic, environmental and structural factors, indicating the need for tailored germination strategies for conservation and agricultural applications. Populations' responses to dormancy release were influenced by their geographic origin and local environmental conditions. Algerian and Italian populations, characterised by broad $T_{10}$ ranges (35-110 hours and 30-120 hours, respectively) and low baseline germination (5-60% and 5-50%, respectively), showed higher dormancy levels likely adapted to arid and Mediterranean conditions. In contrast, French populations demonstrated higher germination, up to 80%, and shorter $T_{10}$ (25-70 hours).

Histological features obtained in the present study are consistent with a balanced dormancy mechanism, potentially allowing flexible germination in response to moisture variability. These variations suggest different dormancy strategies shaped by local climates, with populations from semi-arid regions promoting stronger dormancy with both embryo and seed coat-imposed dormancy as an adaptive strategy [Scialabba *et al*., 2003].

Cold plasma treatment has shown promising results for breaking dormancy in several species [Groot *et al*., 2018], [Dufour *et al*., 2021], [August *et al*., 2023]. In the present study, it notably increased germination, from a mean of 18% in the control group to a mean of approximately 60% for treated seeds. However, the variability in its efficiency highlights the influence of factors such as geographic origin and inherent dormancy mechanisms. Populations with embryo dormancy exhibited notably strong responses, whereas others showed more moderate improvements. Additionally, cold plasma treatment increased germination rate, but its effects were not uniform across all populations, with some even showing delayed germination. Finally, emergence did not show a substantial improvement. These mixed outcomes can be partially attributed to the use of a standard treatment protocol (10 min, 6 kV, 500 Hz) that was not optimised in advance due to logistical limitations. Australian research on cotton has shown for instance the impact of plasma treatment duration on seedling emergence [Groot *et al*., 2018]. A recent study on *B. napus* showed that plasma treatment influenced the germination and seedlings through interaction with the soil microbiome [Kalachova *et al.*, 2024]. Therefore, if the objective is to undertake plant breeding in *Brassica* species, it would be highly advisable to refine the parameters (exposure duration, gap distance, voltage and plasma current) governing the cold plasma properties (electric field strength, reactive species) while controlling seed water content and plasma gas temperature during the treatment.

Comparative analysis of cold plasma with GA3 and scarification treatments provided insights into the selectivity and efficiency of these methods in dormancy release. Cold plasma and GA3 both significantly enhanced germination, achieving mean germination percentages of 67 and 71%, respectively, compared to 79% for scarification. While scarification exhibited the most consistent results, with a narrow interquartile range (IQR) of 21%, cold plasma offered a more uniform approach than GA3, as evidenced by a narrower IQR of 26% for plasma-treated seeds compared to 40% for GA3-treated seeds. Interestingly, plasma treatment was generally more effective in alleviating shallow than relatively deep physiological dormancy. These findings suggest that the standard cold plasma treatment needs further optimisation to reliably serve as a non-chemical alternative to hormone-based methods. Notably, in radish, cold plasma treatment induced changes in seed phytohormone content, particularly of ABA and GA [Degutytė-Fomins *et al.*, 2020].

Additionally, SEM analysis confirmed that cold plasma does not cause significant morphological changes to seed coat surfaces, indicating that its dormancy-breaking action is likely due to physicochemical modifications, such as increasing permeability or altering surface chemistry, rather than mechanical abrasion [Adhikari *et al.*, 2020], [Starič *et al.*, 2022]. Research on *B. napus* showed variations in alterations and chemical composition of seed coat depending on the plasma treatment duration [Kalachova *et al.*, 2024]. Cold plasma also offers practical advantages as it is a dry, non-chemical method that avoids the challenges associated with chemical treatments, such as hormone concentration variability, labour-intensive preparation and regulatory concerns. Cold plasma can be applied uniformly to seed batches, potentially providing more consistent dormancy release outcomes in case of deep physiological dormancy. Nevertheless, this new method for breaking dormancy was unfortunately not sufficient to obtain better greenhouse emergence. Studies of heterotrophic growth post-germination could help to measure the impact of cold plasma on seedling establishment.

This variability highlights the need for population-specific approaches to optimize germination, particularly for conservation and breeding programs aiming to enhance genetic diversity. Moreover, addressing additional dormancy mechanisms, such as photoinhibition observed in related species, including turnip weed (*Rapistrum rugosum* (L.) All.) in Australia and certain African leafy vegetables, may further refine the application of cold plasma in overcoming physiological dormancy traits [Motsa *et al*., 2015], [Manalil *et al*., 2018].

To conclude, we have demonstrated that physiological dormancy exists in wild *B. rapa* with different intensity and that this trait could be linked to the native geographic origin of populations. While cold plasma significantly improved both germination percentage and rate in almost all populations, its effect on seedling emergence was limited, highlighting the need for further refinement of treatment parameters. Compared to traditional methods such as gibberellic acid and scarification, cold plasma offers a sustainable, non-chemical alternative with distinct advantages in addressing physiological dormancy. However, its application must be tailored to population-specific dormancy traits to achieve consistent results. These findings underline the complexity of dormancy regulation in *B. rapa* and the importance of optimising germination protocols for conservation, breeding, and agricultural purposes. Future research should focus on enhancing the efficacy of plasma treatments, exploring their underlying biochemical mechanisms, and integrating them into broader strategies for sustainable seed management and breeding.

# V. Acknowledgements

BrasExplor project (ANR-19-P026-0010-03) has been funded by the PRIMA programme supported by Horizon 2020 (project 1425). Authors are grateful to Valérie Blouin for scarifying many single seeds, Aurore Philibert for her kind and helpful assistance on statistical analyses with R and A. Manseri from the Research Center in Semi-conductors Technology for Energetic (CRTSE) in Algiers for his technical assistance in SEM imaging.